\documentclass[twocolumn,aps,amsmath,amssymb,prb]{revtex4}
\usepackage{subfigure}
\usepackage{graphicx}

\newcommand{\bc}{\begin{center}}
\newcommand{\ec}{\end{center}}
\newcommand{\be}{\begin{equation}}
\newcommand{\ee}{\end{equation}}
\newcommand{\ba}{\begin{array}}
\newcommand{\ea}{\end{array}}
\newcommand{\beq}{\begin{eqnarray}}
\newcommand{\eeq}{\end{eqnarray}}

\begin{document}

\title{Strong disorder renormalization group study of $S=1/2$
  Heisenberg antiferromagnet layers/bilayers with bond randomness,
  site dilution and dimer dilution}

\author{
Yu-Cheng~Lin and Heiko Rieger
}
\affiliation{
Theoretische Physik, Universit\"at des Saarlandes, 66041 Saarbr\"ucken, Germany}

\author{Nicolas Laflorencie}
\affiliation{Department of Physics \& Astronomy, University of British Columbia,
Vancouver, B.C., Canada, V6T 1Z1}

\author{Ferenc Igl\'oi}
\affiliation{
Research Institute for Solid State Physics and Optics, H-1525 Budapest, Hungary\\
Institute of Theoretical Physics, Szeged University, H-6720 Szeged, Hungary}
\begin{abstract}
  Using a numerical implementation of strong disorder renormalization
  group, we study the low-energy, long-distance properties of layers
  and bilayers of $S = 1/2$ Heisenberg antiferromagnets with different
  type of disorder: bond randomness, site and dimer dilution.
  Generally the systems exhibit an ordered and a disordered phase
  separated by a phase boundary on which the static critical exponents
  appear to be independent of bond randomness in the strong disorder
  regime, while the dynamical exponent is a continuous function of the
  bond disorder strength.  The low-energy fixed points of the
  off-critical phases are affected by the actual form of the disorder,
  and the disorder induced dynamical exponent depends on the disorder
  strength. As the strength of bond disorder is increased, there is a
  set of crossovers in the properties of the low-energy singularities.
  For weak disorder quantum fluctuations play the dominant role.
  For intermediate disorder non-localized disorder fluctuations are
  relevant, which become localized for even stronger bond disorder. 
  We also present some quantum Monte Carlo simulations results
  to support the strong disorder renormalization approach.

\end{abstract}
\maketitle

\section{Introduction}

The two-dimensional ($2d$) spin-$1/2$ Heisenberg antiferromagnet has
attracted abiding interest in recent years, mainly motivated by its
relation to high-temperature superconductivity.\cite{fradkin}
According to the Mermin-Wagner theorem,\cite{MERMIN} the N\'eel
antiferromagnetic (AF) long-range order in $2d$ can exist only at zero
temperature, but even then it can still be reduced by quantum
fluctuations. It has been established that at $T=0$ the AF order
survives for several lattices, such as for the square lattice.  The
ordered ground state is accompanied by gapless low-energy excitations,
which, according to spin-wave theory\cite{SPINWAVE} and the non-linear
$\sigma$-model description\cite{SIGMA}, behave as:
\be
\Delta E_{q}\sim L^{-z_{q}},\quad z_q=2,
\label{e_q}
\ee
where $L$ is the linear size of the system, $z_q$ is the dynamical
exponent and the subscript $q$ refers to quantum fluctuations.  The AF
order in the ground state can be suppressed by introducing frustration
(e.g. with diagonal couplings in the square lattice: $J_1-J_2$
model),\cite{FRUSTRAT} by dimerizing the lattice,\cite{2dDIMER} or by
coupling two square lattices to form a
bilayer.\cite{HIDA,MILLIS,SANDVIK-BI} By increasing these disordering
effects, the AF order is reduced progressively and will disappear at
an order-disorder quantum phase transition point.

In real materials impurities and other types of quenched disorder are
inevitably present or can be controlled by doping.  Fluctuations due
to quenched disorder can further destabilize the AF order, resulting
in disordered ground states and random quantum critical points.
Quasi-two-dimensional materials, such as La$_2$CuO$_4$ doped with Mg
(or Zn) and K$_2$CuF$_4$ (or K$_2$MnF$_4$) doped with Mg can be
approximately described by the $2d$ AF Heisenberg model with static
nonmagnetic impurities. In these systems a disorder induced quantum
phase transition from N\'eel order to a disordered spin liquid phase
was observed.\cite{EXP}

Theoretical investigations on the disorder effects in $2d$ Heisenberg
antiferromagnets have been mainly restricted to dilution effects.
Quantum Monte Carlo (QMC) simulations of the diluted square lattice
model showed that the AF long-range order persists up to the classical
percolation point and the critical exponents are identical to those of
classical percolation for all $S$.\cite{SANDVIK} In studies of the
square lattice model with staggered dimers and dimer dilution, unusual
critical properties were found, among others, at the classical (bond)
percolation point there is a critical line with varying
exponents.\cite{SANDVIK_pr} In the $2d$ bilayer Heisenberg
antiferromagnet the random dimer dilution can be introduced by
randomly removing the inter-layer bonds. In recent QMC
simulations,\cite{SANDVIK-DI,VAJK,VOJTA} random quantum critical
points with an universal dynamical exponent $z \approx 1.3$ were
deduced by varying the ratio of the inter-layer and intra-layer
couplings below the percolation threshold.

In the presence of bond randomness, the low-energy properties of the
above mentioned $2d$ random models can be studied by a strong disorder
renormalization group (RG) approach,\cite{RGREV} which was originally
introduced by Ma, Dasgupta and Hu\cite{MDH} for the $1d$ random AF
Heisenberg model.  In a detailed analysis of this RG procedure
Fisher\cite{FISHER1} solved the RG equations for the $1d$ model
analytically and showed that during renormalization the distribution
of the couplings broadens without limit, indicating that the RG flow
goes to an infinite-randomness fixed point.\cite{IRFP} Due to infinite
randomness, approximations in the RG procedure are negligible and the
scaling behavior of the system - both in dynamical and static sense -
is asymptotically exact. The ground state of the $1d$ model, the
so-called random singlet state,\cite{FISHER} consists of effective
singlet pairs and the two spins in a given singlet pair can be
arbitrarily far from each other. Renormalization of the $1d$ model
with enforced dimerization (with different probability distributions
of the even and odd couplings) leads to a random dimer
phase,\cite{HYMAN} which is a prototype of a quantum Griffiths phase.
The singular properties of the Griffiths phase are controlled by a
line of strong disorder fixed points; along this line, the disorder
induced dynamical exponent $z$ varies continuously with the strength
of dimerization. The dynamical exponent, calculated by the RG method,
is presumably asymptotically exact, however the static behavior, such
as the density profiles, are correct only up to the correlation length
in the system.

Variants of the strong disorder RG method have been applied for
various $1d$ and quasi-$1d$ (spin ladders) random Heisenberg models.
In Heisenberg models with mixed ferro- and antiferromagnetic
couplings\cite{WESTERBERG}, during renormalization large spins are
formed and the dynamical properties of these large-spin phases are
different from the Griffith phases, for example the uniform magnetic
susceptibility has a Curie-like low-temperature behavior.  The strong
disorder RG method for more complicated geometries, such as in $2d$,
can only be implemented numerically and the calculated dynamical
exponent $z$ is presumably approximative. However we expect that the
qualitative form of the low-energy singularities is correctly
predicted by these investigations. In previous studies\cite{HAF} $2d$
and $3d$ Heisenberg antiferromagnets with/without frustration in the
presence of bond disorder were numerically studied for random coupling
constants taken from the Gaussian or from the box-like distributions.
In contrast to the $1d$ case, no infinite disorder fixed point is
observed. Non-frustrated models are shown to have a conventional
Griffiths-like random fixed point, whereas the dynamical singularities
of frustrated models are controlled by large-spin fixed points.

In the present paper we extend previous investigations of $2d$ random
Heisenberg models in different directions. First, we consider strong
disorder represented by power-law distribution of the couplings and
study systematically the variation of the dynamical singularities with
the strength of bond disorder. In particular, we are interested in the
localization properties of the low-energy excitations. Second, we
consider non-magnetic impurities and study the combined effect of bond
disorder and site dilution. Our third direction of study considers AF
bilayers with bond disorder and randomly removed inter-layer dimers.
Evidently, with vanishing inter-layer coupling this problem reduces to
our second model.

The paper is organized as follows. The models under investigation as
well as their basic properties are presented in Sec.~\ref{sec:models}.
The strong disorder RG method and the properties of the basic fixed
points are shown in Sec.~\ref{sec:RG}. A description of the QMC stochastic
series expansion method, which is used to support the strong disorder
RG approach, is given in Sec.~\ref{sec:QMC}. Results on the critical 
properties as well as the Griffiths singularities of different
disordered Heisenberg AF models are presented in
Sec.~\ref{sec:results} and discussed in Sec.~\ref{sec:disc}.

\section{Models and phase diagrams}
\label{sec:models}

\begin{figure}
 {\par\centering \resizebox*{64mm}{!}
   {\includegraphics{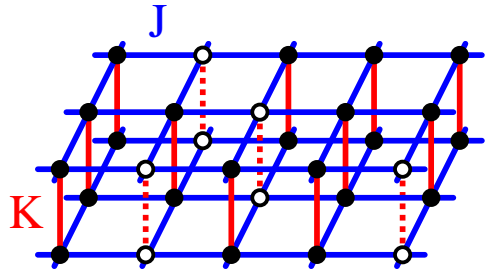}} \par}
 \caption{
 \label{fig:model}
 The diluted bilayer model. Solid circles represent spins and open
 circles indicate the removed dimers. Neighboring spins in each plane
 interact with the coupling $J$, and the interplane coupling is $K$.
}
\end{figure}

\begin{figure}
{\par\centering \resizebox*{84mm}{!}
   {\includegraphics{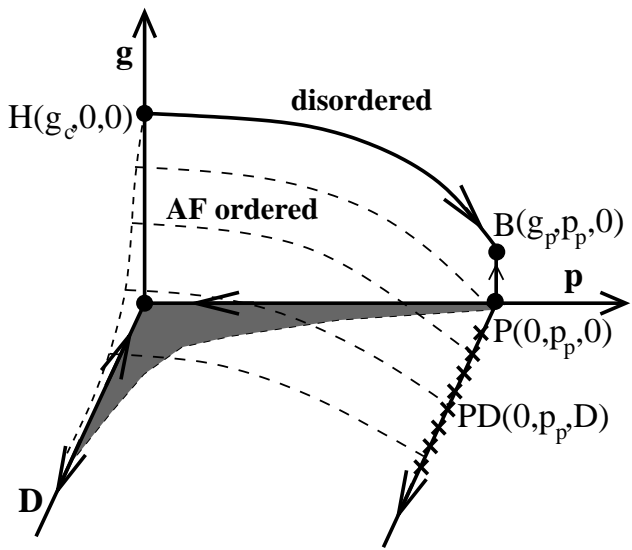}} \par}
 \caption{
\label{fig:phase}
Schematic phase diagram of the dimer diluted bilayer Heisenberg
antiferromagnet, as function of the coupling ratio $g$, the fraction
of the removed inter-plane dimers $p$, and the strength of bond
disorder $D$. The disordered phase and the AF ordered phase are
separated by a critical surface, indicated by dashed lines, which is
located at $p \le p_p$ and $g_c(p,D)$, where $p_p$ is the
site-percolation threshold.  In the model without bond disorder,
$D=0$, there are two unstable fixed points, H and P, as well as a
stable bilayer fixed point B. In the diluted single layer $g=0$ with
bond disorder, the phase boundary is located at the percolation
threshold with universal static and strong disorder dependent
dynamical critical exponents, indicated by the line of fixed points
PD.  In the AF ordered phase the dynamical exponent $z$ is determined
by quantum fluctuations for weak disorder (indicated by a grey region
at $g=0$), whereas $z$ is $D$ dependent for strong disorder.  }
\end{figure}

We start with the definition of the most general model considered in
this paper: the double-layer Heisenberg antiferromagnet with random
dimer dilution (see Fig.~\ref{fig:model}) which is described by the
Hamiltonian:

\beq \mathcal{H}= \sum_{n=1,2}\sum_{\langle i,j\rangle} J_{i,j}\, \epsilon_i
\epsilon_j \mathbf{S}_{i,n}\cdot\mathbf{S}_{j,n}+ \sum_i K_i\,
\epsilon_i\,\mathbf{S}_{i,1}\cdot\mathbf{S}_{i,2}\;.
\label{eq:hamiltonian} \eeq Here $\mathbf{S}_{i,n}$ is a spin-$1/2$
operator at site $i$ of the $n$-th square lattice layer.  The
antiferromagnetic planar (inter-layer) coupling constants $J_{i,j}$
($K_i$) are independently and identically distributed random
variables. The dimer dilution at site $i$ is represented by the
variable $\epsilon_i$, which is $\epsilon_i=0$ with probability $p$
and $\epsilon_i=1$ with probability $1-p$.

To our knowledge, this model has so far only been studied without bond
disorder, i.e. $K_i\equiv K \;\forall i$ and $J_{i,j} \equiv J
\;\forall i, j$.  The schematic phase diagram of this model at zero
temperature in terms of the coupling ratio $g \equiv K/J$ and dilution
$p$ is shown in the plane $D=0$ in Fig.~\ref{fig:phase}. The point at
($g=0$, $p=0$) corresponds to two uncoupled non-diluted square lattice
AF Heisenberg model and exhibits AF long-range order in its ground
state.\cite{REGER} At $p=0$ a finite inter-plane coupling, $g>0$,
causes a tendency for neighboring spins in the adjacent layers to form
singlets and the AF order is therefore reduced. If the coupling ratio
exceeds some critical value, $g>g_c$, the system will undergo a
quantum phase transition from an AF state to a disordered state. This
$T=0$ order-disorder transition is expected to belong to the
universality class of the $3d$ classical Heisenberg model according to
the $\sigma$-model description by Chakravarty {\it et al}.\cite{SIGMA}
Results of recent QMC simulations are in accordance with this
conjecture and the critical ratio is calculated as: $g_c \approx
2.5220$.\cite{SANDVIK-BI,SANDVIK-NEW}

Along the horizontal axis of Fig.~\ref{fig:phase}, i.e. with $g=0$
(and $D=0$) we have two uncoupled site diluted Heisenberg AF planes.
Increasing dilution suppresses AF order progressively and according to
QMC results the quantum phase transition takes place at the classical
site-percolation threshold,\cite{QM-P} $p_p=0.407$. Furthermore, the
critical exponents are those of the classical percolation
transition.\cite{SANDVIK-DI} Now having both dilution, $p>0$, and
finite inter-layer coupling, $g>0$, the phase boundary $g_c(p)$ is
monotonously decreasing with increasing dilution.\cite{SANDVIK-DI,
  VAJK} However even at the percolation threshold there is a finite
critical coupling: $g_c(p_p) \equiv g_p\approx 0.16$, and this
fixed-point, marked by B in Fig.~\ref{fig:phase}, is found to control
the phase transition between the ordered and disordered phases for
$p>0$ and $g>0$.\cite{SANDVIK-DI, VAJK} This fixed point is a
conventional random fixed point with power-law dynamical scaling and
universal exponents.\cite{VOJTA}

In this paper we extend the space of parameters by introducing bond
disorder, such that the intra-layer and inter-layer couplings are
independent and identically distributed random variables taken from
the distributions:
\beq
\pi(J) & = &\frac{J_{\textrm{max}}^{-1/D}}{D}\, J^{-1+1/D},\qquad  \textrm{for } 0<J\le J_{\textrm{max}}; \label{eq:pow} \\
\rho(K)& = &\frac{K_{\textrm{max}}^{-1/D}}{D}\, K^{-1+1/D}, \qquad
\textrm{for } 0<K\le K_{\textrm{max}} \nonumber\;, \eeq 
respectively. Here $D^2 =\textrm{var}[ \ln J]=\textrm{var}[ \ln K]$
measures the strength of disorder ($ {\rm var}[x]$ stands for the
variance of $x$) and the control parameter is defined as
$g=K_{\textrm{max}}/J_{\textrm{max}}$.  Note that an uniform
distribution corresponds to $D=1$.  In particular we are interested in
the properties of the phase diagram, the singularities at the phase
transitions as well as the form of disorder induced low-energy
excitations in the different regions.

\section{The strong disorder RG method and its fixed points}
\label{sec:RG}

The strong disorder RG method\cite{RGREV} is an important tool to
study random quantum systems.  Here we recapitulate the basic
ingredients of the method used for the $2d$ random Heisenberg
antiferromagnet.

The RG proceeds by eliminating at each step a term in the Hamiltonian
with the largest gap separating the ground state and the first excited
state. This decimation process generates new effective couplings
between the remaining sites which are calculated perturbatively.  For
a lattice with more complex structure than a single chain, such as the
bilayer antiferromagnet, the renormalized Hamiltonian contains
effective spins of arbitrary size with a complicated correlated
network and has both antiferromagnetic and ferromagnetic (F)
couplings.  The RG procedure for this Hamiltonian thus consists of two
types of decimation rules, one for singlet formation (for equal-size
spins with an AF bond), and one for cluster formation (for all other
cases). Further details of the RG procedure can be found in
Ref.~[\onlinecite{WESTERBERG,LADDER-M,HAF}].

During renormalization the energy scale $\Omega$, which is set by the
cutoff the energy gaps of the effective Hamiltonian, is gradually
decreasing. In the vicinity of the low-energy fixed point $\Omega^*
\to 0$, the low-energy tail of the distribution of the gaps for a
large finite system of linear size $L$ follows the relation:
\be
P(\Delta,\Omega,L)=L^z\tilde{P}\left(\frac{\Delta}{\Omega},\frac{L^{-z}}{\Omega}\right) \sim L^z 
\left(\frac{\Delta}{\Omega}\right)^{\omega} \sim L^{z(1+\omega)} \Delta^{\omega}\;.
\label{eq:distr}
\ee
which defines the gap exponent $\omega$. The energy-scale and the
length-scale is related by $\Omega \sim L^{-z}$ with the disorder
induced dynamical exponent $z$. Note that with the initial power-law
distribution of the couplings in Eq.(\ref{eq:pow}) the initial gap
exponent is given by $\omega_0=-1+1/D$. At a conventional random fixed
point, we have $\omega/\omega_0=O(1)$, while at a infinite-disorder
fixed point the distribution of the effective gaps broadens without
limit, indicating $\omega/\omega_0 \to \infty$.  If the low-energy
excitations are localized, than the gap distribution for a fixed
$\Delta$ is proportional to the volume of the system:
$P(\Delta,\Omega,L) \sim L^d$. From Eq.(\ref{eq:distr}), we obtain in
this case:
\be
z=z' \equiv \frac{d}{1+\omega}\;,
\label{eq:z_omega}
\ee
here an exponent $z'$ is defined.  Note that at an infinite-disorder
fixed point the dynamical exponent $z$ is formally infinite.

Another characteristic feature of the fixed point is the typical size
of the effective cluster moment, $S_\textrm{eff}=|\sum_i \pm S_i|$,
which is determined by the classical correlation of the spins in the
ground state, and the positive (negative) sign corresponds to an F
(AF) coupling.  $S_\textrm{eff}$ is expected to scale as
$S_\textrm{eff} \sim L^{d \zeta}$.  There are two types of fixed
points concerning the value of $\zeta$: In some models the decimated
spin pairs are typically singlets or the size of the effective spins
has a saturated value, which yields $\zeta=0$ in the low-energy limit;
In some models, mainly with frustration, large effective spins are
formed and if ferromagnetic and antiferromagnetic couplings are
decimated uncorrelated one obtains\cite{WESTERBERG} $\zeta = 1/2$.
This state is called the large-spin phase.

In the RG method static correlations can be measured by considering
the staggered ground-state correlation function $C(r)$ between two
spins at distance $r$.  This is defined as \beq C(r) \equiv
C_{ij}=\langle \eta_{ij}\mathbf{S}_i\cdot \mathbf{S}_j \rangle.  \eeq
where $\eta_{ij}=(-1)^{x_i + y_i+x_j+y_j}$, and $r$ is measured by
1-norm distance (also known as Manhattan distance): $r=r_{ij}\equiv
|x_i -x_j| +|y_i-y_j|$. This choice was made for computational
convenience; in the limit $r\to \infty$ it yields the same asymptotic
behavior of $C(r)$ as the one calculated with the Euclidean distance
$r=\sqrt{(x_i-x_j)^2+(y_i-y_j)^2}$.  In our RG scheme, the
correlations of spin pairs which form an effective spin at each RG
stage, are calculated by \beq \langle \mathbf{S}_i\cdot \mathbf{S}_j
\rangle= \alpha_{ik} \alpha_{jl} \langle
\mathbf{S}_k^{\textrm{eff}}\cdot \mathbf{S}_l^{\textrm{eff}} \rangle,
\eeq where $\alpha_{ik(jl)}=\langle \mathbf{S}_{i(j)} \cdot
\mathbf{S}_{k(l)}^{\textrm{eff}}\rangle/\langle{\mathbf{S}_{k(l)}^{\textrm{eff}}}^2\rangle$
are the proportionality coefficients for each spin. We assume zero
correlation between two spins that do not form an effective spin.
After accumulating the correlations between all decimated spin pairs,
we divide the correlation for a given distance $r$ by $2rL^2$, which
corresponds to the number of pairs a 1-norm distance $r$ apart.  Here
we note that the RG results for static correlations are expected to be
valid only in the vicinity of a (static) critical point. Thus the
calculated correlation functions for the $2d$ problem are
asymptotically correct only in the vicinity of the phase boundary.

Within the RG study, thermodynamics can be understood by stopping the
RG procedure when the energy scale, i.e. the cutoff of energy gaps
$\Omega$ in our case, reaches the thermal energy at a given
temperature $T$.\cite{FISHER,FISHER1} At this scale, almost all
decimated spins are effectively frozen, while almost all remaining
spins involve couplings which are much less than $T$ and hence can be
regarded as free.  The magnetic susceptibility per spin is then mainly
given by the Curie-contribution of those remaining spins and is given
by
\beq \chi(T) \sim \frac{1}{TL^d} \sum_i^{n_T} S_i(S_i+1)\;,
\eeq
where the summation runs over all clusters left at the given
temperature $T$, and $S_i$ is the (effective) spin moment. In the
low-temperature limit the susceptibility generally behaves as a
power-law:
\be
\chi(T) \sim T^{-\theta}
\label{eq:theta}
\ee
If during renormalization there is no large-spin formation i.e.
$\zeta=0$, then $\theta=\omega$ in the low $T$ limit, whereas in the
large-spin phase with $\zeta=1/2$ there is a Curie-like dependence:
$\theta=1$. Singularities of the specific heat or the magnetization
can be calculated similarly, see Ref.[\onlinecite{RGREV}].

\section{Quantum Monte Carlo method}
\label{sec:QMC}
\subsection{Description of the method}
Here we use the QMC stochastic series expansion (SSE) method within a directed
loop framework introduced by Syljuasen and Sandvik in Ref.~\onlinecite{SANDVIK-LOOP}.
Starting with a general Heisenberg Hamiltonian with random exchanges $J(b)$, we can
rewrite it as a sum over diagonal and off-diagonal operators

\be{\mathcal{H}}=-\sum_{b=1}^{N_b}
J(b)\Bigl[H_{1,b}-H_{2,b}\Bigr]\ee where $b$ denotes a bond
connecting a pair of interacting spins $(i(b),j(b))$, and $N_b$ is the 
total number of bonds.
\be
H_{1,b}=C-S^{z}_{i(b)}S^{z}_{j(b)}
\ee
is the diagonal part and the off-diagonal part is given by
\be
H_{1,b}=\frac{1}{2}[S^{+}_{i(b)}S^{-}_{j(b)}+S^{-}_{i(b)}S^{+}_{j(b)}],
\ee
in the basis
$\{|\alpha\rangle\}=\{|S^{z}_{1},S^{z}_{2},...,S^{z}_{L}\rangle\}$. 
The constant
$C$ which has been added to the diagonal part ensures that all
non-vanishing matrix elements are positive.  The SSE algorithm
consists in Taylor expanding the partition function
$Z={\rm{Tr}}\{e^{-\beta{\mathcal{H}}}\}$ up to a cutoff
${\mathcal{M}}$ which is adapted during the simulations in order to
ensure that all the elements of order higher than ${\mathcal{M}}$ in
the expansion do not contribute. So
\be
Z=\sum_{\alpha}\sum_{S_{\mathcal{M}}}
\frac{\beta^{n}({\mathcal{M}}-n)!}{{\mathcal{M}}!}\Bigl\langle\alpha\Bigl|
\prod_{i=1}^{\mathcal{M}}J(b_i)H_{a_i,b_i}
\Bigr|\alpha\Bigr\rangle,
\ee
where $S_{\mathcal{M}}$ denotes a sequence of operator indices 
\be
\label{Op.Ind}
S_{\mathcal{M}}=[a_1,b_1],[a_2,b_2],...[a_{\mathcal{M}},b_{\mathcal{M}}]
\ee
with $a_i=1,2$ corresponds to the type of operator (diagonal or not) and 
$b_i=1,2,...N_b$ is the bond index. 
A Monte Carlo configuration is therefore defined by a state 
$|\alpha\rangle$ and a sequence $S_{\mathcal{M}}$.
Of course, a given operator string does not contain 
$\mathcal{M}$ operators of type $1$ or $2$, but only $n$; so 
in order to keep constant the size of $S_{\mathcal{M}}$, ${\mathcal{M}}-n$ 
unit operators $H_{0,0}=1$ have been inserted in the string, 
taking into account all the possible ways of insertions. 
The starting point of a simulation is given by a random initial 
state $|\alpha\rangle$ and an operator string containing ${\mathcal{M}}$ 
unit operators $[0,0]_1,...,[0,0]_{\mathcal{M}}$. 
The first step is the {\it{diagonal update}} which consists in exchanging 
unit and diagonal operators at each position $p$ $[0,0]_p\leftrightarrow 
[1,b_i]_p$ in $S_{\mathcal{M}}$ with Metropolis acceptance probabilities
\begin{eqnarray}
P_{[0,0]_p\rightarrow [1,b]_p}=
{\rm{min}}(1,\frac{J(b)N_b\beta
\Bigl\langle\alpha(p)\Bigl|H_{1,b}\Bigr|\alpha(p)\Bigr\rangle}{{\mathcal{M}}-n}),\\
P_{[1,b]_p\rightarrow [0,0]_p}=
{\rm{min}}(1,\frac{{\mathcal{M}}-n+1}{J(b)N_b\beta
\Bigl\langle\alpha(p)\Bigl|H_{1,b}\Bigr|\alpha(p)\Bigr\rangle}).
\end{eqnarray}
During the ``propagation'' from $p=1$ to $p={\mathcal {M}}$, the ``propagated'' 
state 
\be
|\alpha(p)\rangle\sim \prod_{i=1}^{p}H_{a_i,b_i}|\alpha\rangle
\ee
is used and the number of non-unit operators $n$ can varies at each index $p$. 
The following step is the {\it{off-diagonal update}}, also called the 
{\it{loop update}}, carried out at $n$ fixed. 
Its purpose is to substitute $[1,b_i]_p\leftrightarrow [2,b_i]_p$ in a 
non-local manner but in a cluster-type update. 
At the $SU(2)$ AF point, the algorithm is deterministic because one can 
build all the loops in a sole way.\cite{SANDVIK-LOOP} 
One MC step is composed by one {\it{diagonal}} and {\it {off-diagonal}} 
updates. Before starting the measurement of physical observables, one 
has to perform equilibration steps, notably necessary to adapt the cutoff 
${\mathcal{M}}$. 


\subsection{Monte Carlo measurement issues}

The precise determination of physical observables using QMC suffers
obviously from statistical errors since the number of MC steps is
finite. As we deal with disordered spin systems, the sample to sample
fluctuation is another source of errors. However one can use a relatively small
number of MC steps for each sample (typically $\sim 100$ at each temperature)
since for the strong disorders considered here, 
the sample to sample variation produces larger error
bars than statistical errors. Than we need to perform a disordered samples
average over a significant number of realization: typically we use $10^3$
samples.

In order to study the low temperature properties, we use the 
$\beta$-doubling strategy introduced by Sandvik~\cite{SANDVIK} to accelerate the
cooling of the system during a QMC simulation.
Such a scheme is a very
powerful tool because it allows to reach extremely low temperatures
rather rapidly {\it{and}} reduces considerably equilibration times in
the MC simulation.  The procedure is quite simple to implement and its
basic ingredient consists in carrying out simulations at successive inverse
temperatures $\beta_n=2^n$, $n=0,1,...,n_{max}$. Starting with a given
sample at $n=0$ we perform a small number of equilibration steps
$N_{eq}$ followed by $N_m=2 N_{eq}$ measurement steps.  At the end of
the measurement process, $\beta$ is doubled (i.e. $n \rightarrow n+1$)
and in order to start with an ``almost equilibrated'' MC
configuration, the starting sequence used is the previous
$S_{\mathcal{M}}$ doubled, i.e.,
\be
S_{2\mathcal{M}}=[a_1,b_1],...[a_{\mathcal{M}},b_{\mathcal{M}}][a_{\mathcal{M}},b_{\mathcal{M}}],...,[a_1,b_1].
\ee

\section{Numerical results}
\label{sec:results}

In practice we started with a finite system of linear size $L$ (up to
$L=64$) with periodic boundary conditions for each single layer, and
performed the RG procedure until the last effective spins (or the last
spin singlet).  The static characteristics of the system, in
particular in the vicinity of the phase boundaries, can be deduced
from the average spin-spin correlation function. On the other hand,
the form of the dynamical singularities can be obtained from the
temperature dependence of the uniform susceptibility and from the
distribution of the first energy gaps corresponding to the energy
scale of the last decimation step.  From the histogram of the gaps we
have extracted the gap exponent $\omega$ and the dynamical exponent
$z$, as discussed in Sec. \ref{sec:RG}. Depending on the size of the
system we have considered $1000-10000$ disorder realizations. 

For the single layer we also compare the RG results with
QMC simulations performed at finite temperature 
on $32\times 32$ square lattices and averaged over $1000$ random samples.

In what follows, we present the phase diagram of the system and the
properties of the different bond-randomness driven phase transitions.
The dynamical properties of the ordered and disordered phases are
discussed afterwards.

\subsection{Phase diagram and critical properties}
\label{sec:PD}

Our main results are summarized in the schematic phase diagram of the
system depicted in Fig.~\ref{fig:phase}. It contains two phases: the
ordered AF phase and the disordered paramagnetic phase.  The phase
transition between these two phases is controlled by several fixed
points as shown in the phase diagram.  The fixed points located at
$D=0$, denoted by H, B and P in Fig.~\ref{fig:phase}, had already been
carefully studied by QMC
simulations.\cite{VAJK,SANDVIK-BI,VOJTA,SANDVIK-DI} The measured
critical exponents at these fixed points are shown in
Table~\ref{table:1}, along with the results for $D>0$ obtained from
our study.

\begin{table}
\caption{Critical exponents at the fixed points of the bilayer Heisenberg antiferromagnet with
random dimer dilution and bond disorder, see in Fig.~\ref{fig:phase}.
H: non-random bilayer (classical 3d Heisenberg model); P: diluted single layer (classical 2d
percolation); B: dimer diluted bilayer; PD: diluted single layer with bond disorder.
In the last rows critical exponents measured at two general points of the critical surface are presented.
\label{table:1}}
 \begin{tabular}{|c|c|c|c|c|}  \hline
  fixed point & position $(g,p,D)$& $\beta/\nu$ & $z$ & $\nu$ \\ \hline
  H\cite{XXX} & $(g_c,0,0)$ & $0.51$ & $1$ & $0.70$  \\
  P\cite{STAUFFER} & $(0,p_p,0)$& $5/48$ & $91/48$ & $4/3$ \\
  B\cite{VOJTA} & $(g_p,p_p,0)$& $0.56$ & $1.31$ & $1.16$ \\ 
  PD & $(0,p_p,D>0)$& $0.50$ & $\sim 3.2D$ &  \\ \hline
   & $(1.2,0.33,0.7)$& $0.56$ & $1.36$ &  \\ 
   & $(7.5\cdot10^{-4},0.33,3)$& $0.80$ & $5.13$ &  \\ \hline
  \end{tabular}
  \end{table}

\begin{figure}[!h]
 {\par\centering \resizebox*{76mm}{!}
    {\includegraphics{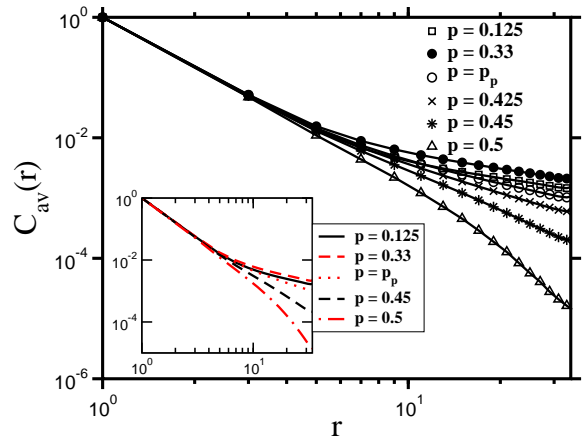}} \par}
 \caption{
 \label{fig:corr_K0}
 Log-log plot of the average spin-spin correlation function at $g=0$
 measured for a $L=64$ lattice with bond randomness $D=3$ and $D=10$
 (inset) for different site dilution $p$. The data are scaled to unity
 at $r=1$.  For $p > p_p$ the curves show downward curvature,
 indicating a faster decay than a power law characteristic to the
 disordered phase. At the percolation threshold $p=p_p$, the decay
 exponent is $2 \beta/\nu \approx 1.01(7)$, which does not depend on
 $D$.  For $p<p_p$ the curves bend upward, indicating a finite
 limiting value and a characteristic of the AF ordered phase.  }
\end{figure}

\begin{figure}[!h]
 {\par\centering \resizebox*{70mm}{!}
    {\includegraphics{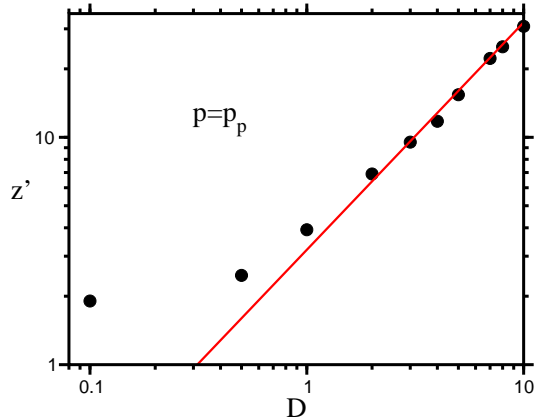}} \par}
 \caption{
 \label{fig:g0_z_pc}
 Disorder dependence of the $z'$ exponent at the percolation threshold
 (at the line of fixed points PD in Table \ref{table:1}) in a log-log
 plot. The slope of the straight line is unity.  }
\end{figure}

  We first consider the fixed points (PD) at the percolation threshold
  $p=p_p$ for $g=0$.  Fig.~\ref{fig:corr_K0} shows the average
  spin-spin correlation function $C_{\textrm{av}}(r)$ at $g=0$ for
  different dilution $p$ and for strong bond randomness,
  $D=3$($D=10$). From $p<p_p$ to $p>p_p$ the decay of
  $C_{\textrm{av}}(r)$ in the log-log plot changes from a upward to a
  downward curvature, and at the percolation threshold a power-law
  decay is found, which implies that the position of the phase
  transition in the site-diluted single layer Heisenberg
  antiferromagnet is robust against strong bond disorder. In
  comparison to the case without bond disorder, the decay of the
  critical average correlation function is, however, faster. From the
  decay of $C_{\textrm{av}}(r=L/2) \sim L^{-2\beta/\nu}$ for different
  system sizes $L$, the decay exponent is found independent of the
  strength of the disorder for $D\ge 3$, and estimated as
  $2\beta/\nu=1.01(7)$, while $2\beta/\nu=0.21$ for $D=0$.  This
  indicates that the percolating cluster is no longer ordered in the
  presence of strong bond randomness.\cite{YRH} Unlike the decay
  exponent of $C_{\textrm{av}}(r)$, which is $D$-independent, the
  dynamical exponent $z'$ obtained from the slope of the gap
  distribution is found to depend linearly on the strength of the
  disorder in the large $D$ region: $z \approx 3.2 D$, as shown in
  Fig.~\ref{fig:g0_z_pc}. For weak bond disorder $D<1$, instead, we
  find that $z'$ approaches to the value $z=91/48$ for the $D=0$ case.

\begin{figure}
 {\par\centering \resizebox*{84mm}{!}
    {\includegraphics{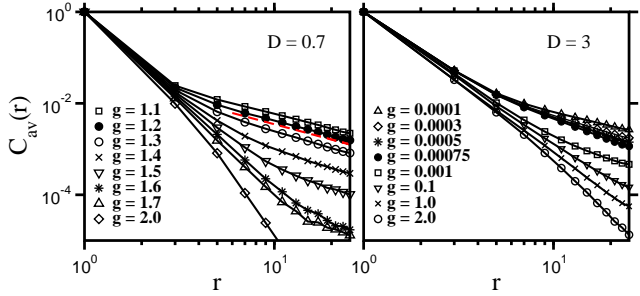}} \par}
 \caption{
 \label{fig:corr_K}
The in-plane average spin-spin correlations of the double-layer AF model versus $r$ in log-log plots 
for bond randomness $D=0.7$ (left) and $D=3$ (right) for different values of the bilayer coupling at  $p=0.33$ for $L=48$.
The data are scaled to unity at $r=1$. 
We observe a crossover from a upward curvature through a power-law decay to a downward curvature. 
The order-disorder transition point $g_c\approx 1.2$ shows an asymptotically linear dependence
in the large $r$ regime with a slope $2 \beta/\nu \approx 1.12(4)$ (indicated by a broken line) which
is approximately the same as for $g=0$. For $D=3$ the transition point shifts to a very small value of $g_c=0.00075$
and the critical exponent is estimated as $2\beta/\nu \approx 1.60(4)$.
}
\end{figure}

\begin{figure}[!h]
 {\par\centering \resizebox*{84mm}{!}
    {\includegraphics{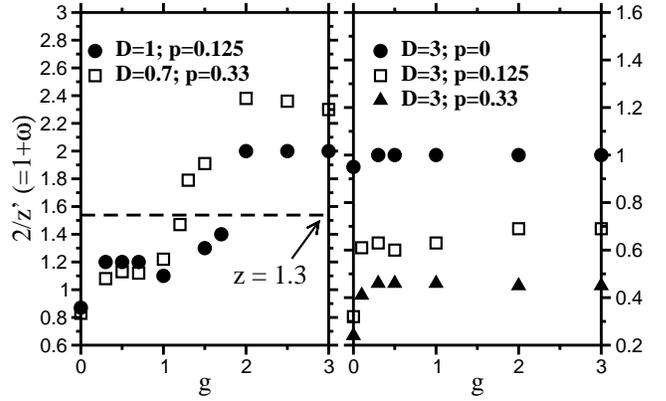}} \par}
 \caption{
 \label{fig:2dbi-gap}
 Variation of the gap exponent with the coupling ratio $g$ for weaker
 (left) and stronger (right) bond disorder for different values of the
 dimer dilution. Note that in the ordered phase $g < g_c$ as well as
 in the disordered phase $g > g_c$, $z'$ is approximately independent
 of $g$.  For weaker bond disorder there is a jump at the transition
 point $g=g_c$ and the dynamical exponent is close to the value
 $z(g_c) \approx 1.3$ at the fixed point B for $D=0$ case, which is
 denoted by a broken line.  For strong disorder (right) the transition
 point is located at a very small $g$ so that it cannot identified in
 the figure.  }
\end{figure}

Now we turn to the phase boundary for finite bilayer coupling $g>0$.
At a given $p<p_p$ and a fixed $D$, we calculated the average
spin-spin correlations $C_{\textrm{av}}(r)$ for different values of
the bilayer coupling $g$. As illustrated in Fig.~\ref{fig:corr_K} for
$p=0.33$, we find that the decay behavior of $C_{\textrm{av}}(r)$
changes its characteristic from the one for the AF ordered phase to
the one for the disordered phase as a critical value of $g_c$ is
traversed. For weak bond disorder $D=0.7$, the critical coupling is
located around $g_c=1.2$ and we note that the decay exponent of the
critical correlation, $2\beta/\nu \approx 1.12$, is approximately the
same as for $D=0$. For strong bond disorder $D=3$, the phase boundary
shifts to a very small value of $g_c\approx 0.00075$ with the critical
exponent $2\beta/\nu \approx 1.6$. The extreme small value of $g_c$,
which decreases even with $D$, makes the investigation on
$D$-dependence of the critical exponents difficult.  From our results
for $C_{\textrm{av}}(r)$ up to $D=5$, the decay exponent $\beta/\nu$
appears to be $D$-independent for a given $p$ in the strong disorder
regime, while it varies with the dilution $p$.  To locate the critical
bilayer coupling $g_c$ we also made use of the results for the
dynamical exponent $z'$, cf. Fig.~\ref{fig:2dbi-gap} for
$p=0,\,p=0.125$ and $0.33$. As $g$ is increased, the dynamical
exponent is approximately independent of the value of $g$, but jumps
to another $g$-independent value around the transition point.  For
weak bond disorder, such as $D=0.7$ for $p=0.33$, we find $z'\approx
1.36$, which is close to the value found for the case without bond
disorder\cite{VOJTA}.  For strong disorder, in which case the RG
approach is expected to be more appropriate, the dynamical exponent
increases with $D$, which is a tendency already noticed for $g=0$.

To summarize our numerical findings indicate two different regimes of
phase transition.  For {\it weak bond disorder} the static critical
exponent $\beta/\nu$ as well as the dynamical exponent $z'$ seems to
coincide with the values for the case without bond disorder. For {\it
  strong bond disorder} the critical coupling $g_c$ is reduced to a
very small value, the static exponent approaches a $D$ independent,
but dilution dependent value, whereas the dynamical exponent at the
transition point depends (linearly) on the strength of the bond
randomness.  The position of the order-disorder transition for a
single layer (corresponding to $g=0$) is located at the percolation
threshold. Along the line of PD fixed points, the exponent $\beta/\nu$
deviates from the value for $D=0$, but seems to be $D$-independent,
while the dynamical exponent exhibits a linear dependence on $D$ in
the large $D$ limit.

\subsection{Griffiths singularities in the ordered phase}

As discussed in the preceding subsection, the random dimer diluted
bilayer antiferromagnet exhibits AF order, provided $p<p_p$ and the
bilayer coupling is sufficiently small.  The low-energy fixed points
governing the Griffiths singularities in the ordered phase are of
different types in the specific regions. These fixed points are in
turn an effective singlet for $p=0$ and $g=0$ (single layer without
site dilution), a large-spin fixed point for $0<p<p_p$ and $g=0$
(single layer with site dilution), and an effective singlet for
$0<p<p_c$ and $0<g<g_c$ (bilayer with dimer dilution). In the
following we study these different cases separately.

\subsubsection{Two-dimensional undoped antiferromagnet}
\label{sec:undoped}

\begin{figure}[!h]
 {\par\centering \resizebox*{76mm}{!}
    {\includegraphics{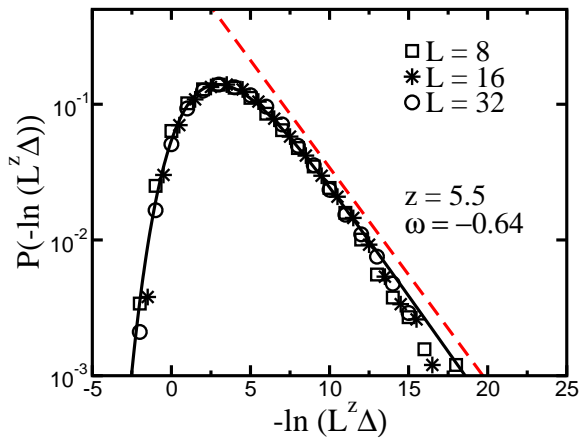}} \par}
 \caption{
 \label{fig:2dhaf_gap}
 A scaling plot of the log-energy gaps for the $2d$ antiferromagnet
 with strong bond randomness $D=8$ obtained from 10000 samples for
 each size. The gap exponent, $\omega \approx -0.64$, follows from the
 slope at small energy gaps and the dynamical exponent, $z \approx
 5.5$, is determined by the fit parameter in Eq.(\ref{eq:distr}). Note
 that the relation in Eq.(\ref{eq:z_omega}) is satisfied, implying
 that the low energy excitations are localized. The full line
 represents the Fr\'echet distribution given in
 Eq.(\ref{eq:frechet}).}
\end{figure}

We start by discussing the results for the two-dimensional random
Heisenberg model, which corresponds to $g=0$ and $p=0$. A recent
numerical study\cite{LAFLORENCIE} suggested that the AF order in this
region vanishes only in the limit of infinite bond randomness.  In our
preliminary study\cite{HAF} we showed that the low-energy fixed point
of the model is conventional, however, the dependence on the strength
of disorder was not investigated extensively. Here we calculate the
gap exponent $\omega$ and the related exponent $z'$ defined in
Eq.(\ref{eq:z_omega}), as well as the dynamical exponent $z$, as a
function of the disorder strength $D$.  The gap exponent $\omega$ is
obtained from the slope of the distribution of the log-gaps in the
small gap limit, whereas the dynamical exponent is determined from the
optimal scaling collapse of the curves according to
Eq.(\ref{eq:distr}) as illustrated in Fig.~\ref{fig:2dhaf_gap}.  For
localized excitations the scaling curve is conjectured\cite{jli06}
from extreme-value statistics to be described by the Fr\'echet
distribution\cite{galambos} \be \tilde{P}_1(u)=\frac{d}{z} u^{d/z-1}
\exp(-u^{d/z})\;,
\label{eq:frechet}
\ee
with $d=2$ and $u=u_0 L^z\Delta$, where $u_0$ is a non-universal constant. 

\begin{figure}[!h]
 {\par\centering \resizebox*{76mm}{!}
    {\includegraphics{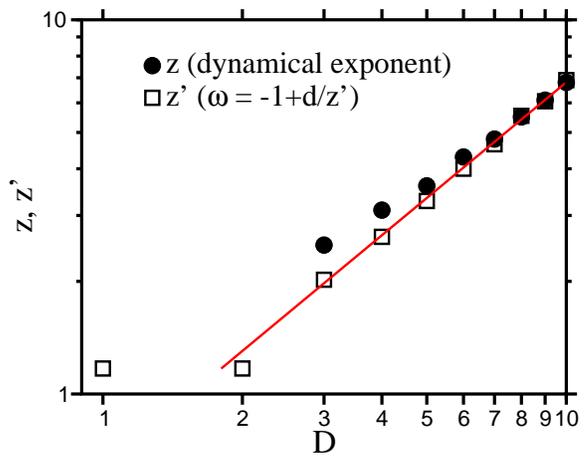}} \par}
 \caption{
 \label{fig:D-omega}
 Variation of the disorder induced dynamical exponent $z$ and the
 exponent $z'$ with the bond randomness strength $D$ at $g=0$ and
 $p=0$ in a log-log plot. Note that the dependence for $D\ge 3$ is
 approximately linear and the values of $z$ and $z'$ fit well for $D
 \ge 7$, indicating that the low-energy excitations are localized.  }
\end{figure}

Both $z$ and $z'$ have an approximately linear $D$-dependence in the
strong disorder region ($D \ge 3$) as shown in Fig.~\ref{fig:D-omega},
while no significant disorder dependence ($\omega \approx 0.7$) is
found for weak disorder.\cite{HAF} The exponents $z$ and $z'$ are
found identical only for quite strong disorder $D \ge 7$.  This
indicates that the low-energy excitations are localized only in the
strong disorder regime.

We note that the vanishing energy gaps calculated by the RG approach
are solely induced by disorder. However, quantum fluctuations also
induce vanishing gaps which are characterized by a dynamical exponent,
$z_q=2$, see in Eq.(\ref{e_q}). The true dynamical exponent is then
given by $z_{\textrm{true}}=\max\{z_q, z\}$, so that
$z_{\textrm{true}}=z_q=2$ for weak randomness $D<3$.

\begin{figure*}
\includegraphics[width=12cm,clip]{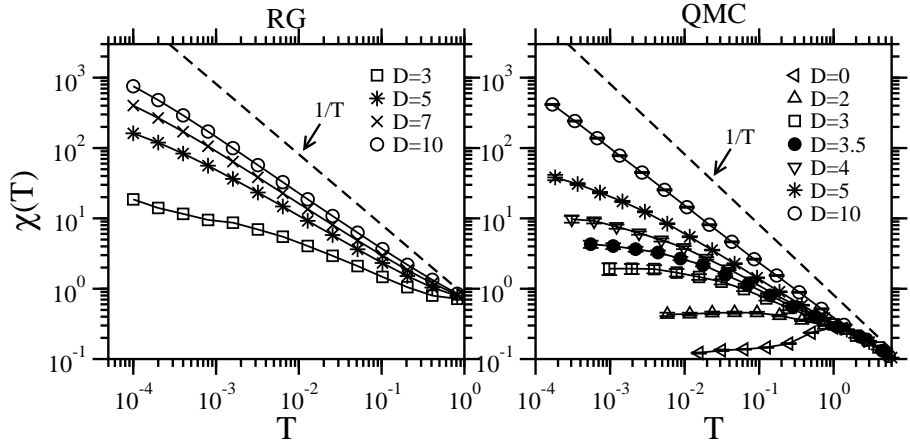}
 \caption{
\label{fig:chi-haf}
 Disorder average uniform susceptibility $\chi(T)$ 
 as a function
 of temperature $T$ for various disorder strength $D$ at $g=p=0$. 
 Left panel: RG results. From
 the low-temperature regime ($T\lesssim 10^{-2}$) the  exponent $\theta$ is
 estimated as: $\theta=0.36$ for $D=3$, $\theta=0.60$ for $D=5$,
 $\theta=0.71$ for $D=7$ and $\theta=0.77$ for $D=10$.  For all
 cases studied, the temperature dependence deviates from Curie-like
 $1/T$ behavior indicated by the dashed line. Right panel: QMC 
 results obtained on systems of $32 \times 32$ spins. The exponent $\theta$ is
estimated in a range of $T\in [T^*,1]$ as: $\theta=0.37$ for $D=3$ ($T^*\simeq
0.02$), $\theta=0.45$ for $D=3.5$ ($T^*\simeq
0.01$), 
$\theta=0.52$ for $D=4$ ($T^*\simeq
0.01$), $\theta=0.61$ for $D=5$ ($T^*\simeq
0.002$) and $\theta=0.81$ for $D=10$ ($T^*\simeq
0.0001$).}
\end{figure*}

We have also calculated the uniform magnetic susceptibility as a
function of the temperature, which is shown in Fig.~\ref{fig:chi-haf}
for different disorder strength.  Both RG and QMC results are shown 
and they display an excellent agreement. 
For strong bond randomness, the
low-$T$ susceptibility exhibits power-law temperature dependence as
given in Eq.(\ref{eq:theta}) and the exponent $\theta$ is disorder
dependent (see table~\ref{table:theta}). The same behavior of 
the magnetic susceptibility has been
found for the antiferromagnetic spin-$1/2$ ladders.\cite{LADDER-Y}
Note however that the QMC results, shown on the right panel of Fig.~\ref{fig:chi-haf}, 
display a slow saturation of $\chi$ when $T\to 0$ (at least for $D\le 5$) which is
not a finite size effect~\cite{NOTE}
but a signature of a tendency towards N\'eel ordering at
$T=0$.\cite{LAFLORENCIE}

{\bf{
\begin{table}
\caption{Exponent $\theta$ of the divergence of the uniform susceptibility for
various disorder strengths $D$ for $p=g=0$.
Comparison between RG and QMC estimates. 
\label{table:theta}}
 \begin{tabular}{c|c|c}  \hline
  $D$ & $\theta_{\rm{RG}}$ &$\theta_{\rm{QMC}}$ \\ \hline
  3& 0.36 & 0.37  \\
  5 & 0.60& 0.61 \\
  10 & 0.77& 0.81 \\ 
  \end{tabular}
  \end{table}
  }}

Finally, we note that effective spins with size larger than $1/2$ are
formed during RG procedure due to the generation of F couplings.  In
the low-energy limit, the overall strength of the F couplings,
however, becomes much weaker than that of the AF couplings, which
leads to the disappearance of large effective spins and the singlet
ground state.  This agrees with the Marshall's theorem\cite{MARSHALL}
which states that the ground state of a bipartite AF Hamiltonian with
equal size sublattices is a total spin singlet.

\subsubsection{Two-dimensional antiferromagnet with site dilution}
\label{sec:diluted}

The low-energy behavior of the site-diluted Heisenberg antiferromagnet
is controlled by a large-spin fixed point, which is different from the
undoped case where the last decimated pair of spins is an effective
singlet. The situation is similar to that of antiferromagnetic
spin-$1/2$ ladders with random site dilution. In this case Sigrist et
al\cite{SIGRIST} argued that if two vacancies are in the same
sublattice, the ground state is no longer a singlet thus there are
effective spins of size larger than $1/2$.  This has been verified by
numerical strong-disorder RG calculations.\cite{LADDER-LS} In the $2d$
site-diluted case we also observed in our numerical RG calculation
that the energy gap associated with an effective F coupling may become
the largest gap to be decimated at some stage of the RG, especially in
the low-energy regime. This will then lead to the formation of large
effective spins as described in Sec.~\ref{sec:RG}.

\begin{figure}[!ht]
 {\par\centering \resizebox*{76mm}{!}
    {\includegraphics{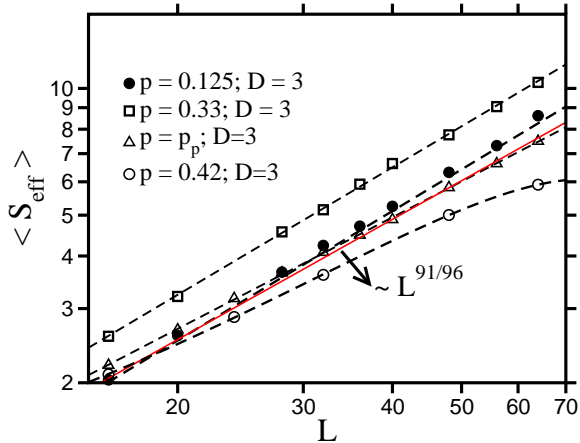}} \par}
 \caption{
 \label{fig:spin}
 Variation of the disorder averaged spin size $\langle
 S_{\textrm{eff}} \rangle$ with the linear system size $L$ in a
 log-log plot for different site dilution $p$ at $D=3.$ for the single
 layer antiferromagnet $g=0$.  For $p<p_p$ the spin size follows
 $\langle S_{\textrm{eff}} \rangle \sim L$, indicated by the broken
 lines, whereas at $p=p_p$ the asymptotic power for large system sizes
 agrees with $91/96$.  }
\end{figure}

We calculated the average size of the effective spin at the last
decimation step $\langle S_{\textrm{eff}} \rangle$ for various
dilution concentrations ($p=0.125,~0.33$ and at $p_p$) and system
sizes $L$.  In the ordered phase, below the percolation threshold, $p
\le p_p$, the average spin size is found to increase linearly with the
system size: \beq \left \langle S_{\textrm{eff}}\right\rangle \sim L,
\label{eq:spin} \eeq which is demonstrated in Fig.~\ref{fig:spin}.
This result agrees with the scenario for the large-spin phase, as
discussed below Eq.(\ref{eq:z_omega}).  At the percolation threshold
the same argument leads to $\langle S_{\textrm{eff}}\rangle \sim
L^{d_f/2}$, with $d_f=91/48$ being the fractal dimension of the
percolation cluster\cite{STAUFFER}.

\begin{figure}[!h]
 {\par\centering \resizebox*{84mm}{!}
    {\includegraphics{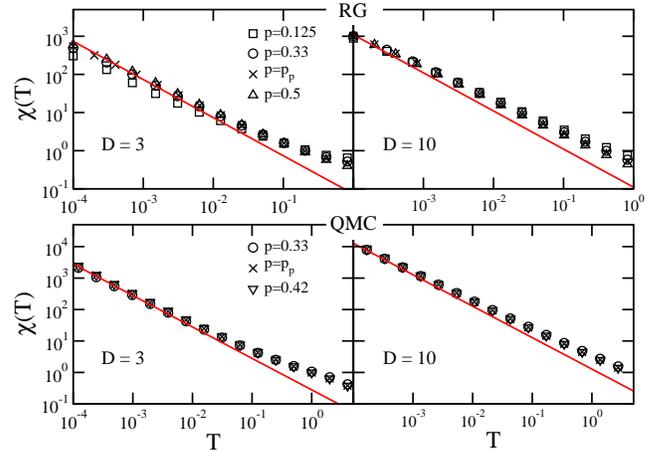}} \par}
 \caption{
 \label{fig:chi-curie}
 Temperature dependence of the uniform susceptibility per size for a
diluted single layer, $g=0$, in
 log-log plots for various dilution concentrations $p$ and for
 different bond random strength $D=3$ (left) and $D=10$ (right).  
 The Curie-like $1/T$ behavior is indicated by straight lines. 
 Both RG (upper panels) and QMC (lower panels) are shown}.
\end{figure}

A hallmark of the large spin phase is the universal temperature
dependence of some thermodynamic quantities, in particular the
disorder averaged uniform susceptibility given in Eq.(\ref{eq:theta})
shows a Curie-like behavior at low temperatures. This is checked in
Fig.~\ref{fig:chi-curie} in which the susceptibility  obtained from both RG and QMC 
is plotted for
different strength of bond randomness and dilution concentrations. For
not too strong bond disorder the agreement with the Curie-law is good,
while for strong bond disorder this agreement is observed only for
very low temperatures. Note that in the undoped regime the
susceptibility exponent $\theta$ is a continuous function of the
disorder, see in Fig.\ref{fig:chi-haf}.

\begin{figure}[!h]
 {\par\centering \resizebox*{76mm}{!}
    {\includegraphics{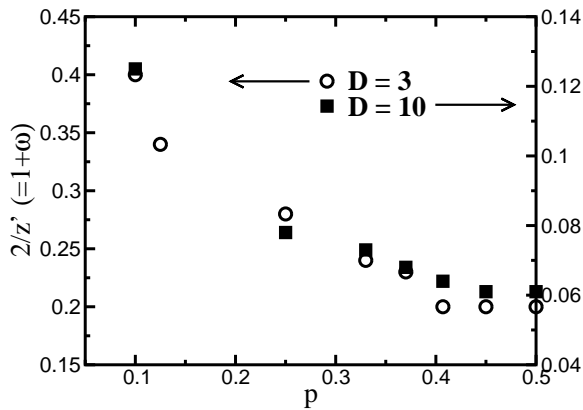}} \par}
 \caption{
 \label{fig:g0_z}
 The gap exponent $\omega$ in the diluted single layer antiferromagnet
 ($g=0$) for different dilution and bond disorder. Note that in the
 disordered phase, above the percolation threshold $p>p_p$, the gap
 exponent is practically independent of the dilution.  }
\end{figure}

From the distributions of the low-lying energy gaps, we obtained the
dynamical exponent $z$ and the gap exponent $\omega$. Unlike for the
undoped model, the exponent $z$ and $z'=2/(1+\omega)$ in general do
not agree with each other for $0<p<p_p$, even in the regime of strong
bond disorder. This indicates that low-energy excitations are not
localized due to the formation of large spins.  Fig.~\ref{fig:g0_z}
presents the exponent $z'$ as a function of $p$ for $D=3$ and $D=10$.
For a given bond disorder, the exponent varies continuously with $p$
in the ordered phase ($p<p_p$), while going approximately to a
constant in the disordered phase ($p>p_p$). We remind that to obtain
the true dynamical exponent $z_{\textrm{true}}$ one should also
consider the effect of quantum fluctuations and thus
$z_{\textrm{true}}=\max\{z_q,z\}$ in the ordered phase.

\subsubsection{The double-layer Heisenberg antiferromagnet}

In the presence of random bilayer couplings $g>0$, the low energy
properties of the ordered phase are controlled by an effective
singlet, both for $p=0$ and $0<p<p_c$, which in turn is the same as
for the single layer undoped model, see in Sec.\ref{sec:undoped}.
Indeed, we observed similar low-energy properties. The dynamical
exponent $z$, and the exponent $z'$ are disorder dependent, but vary
only weakly with the bilayer coupling $g$, see Fig.\ref{fig:2dbi-gap}.
$z$ and $z'$ are identical only for strong enough disorder, when the
low-energy excitations are expected to be localized. The average
uniform susceptibility has a disorder dependent low-temperature
behavior and the exponent $\theta$, corresponds to the gap exponent
$\omega$.

\subsection{Griffiths singularities in the disordered phase}

The disordered phase of the system is divided into two parts with
different low-energy properties:

$\bullet$ Above the percolation threshold $p>p_p$ and $g=0$ the spins
form only finite connected clusters. As a consequence the average
effective spin has a finite value, as shown in Fig.~\ref{fig:spin} for
$p=0.42$. Due to the unpaired spins in the isolated connected spin
clusters the average uniform susceptibility is Curie-like (see
Fig.~\ref{fig:chi-curie} for $p=0.5$).  The dynamical exponent $z$ and
the gap exponent $\omega$ depend approximately linearly on the bond
disorder $D$, they exhibit however no significant dependence on $p$,
as shown in Fig.~\ref{fig:g0_z}.

\begin{figure}
 {\par\centering \resizebox*{76mm}{!}
    {\includegraphics{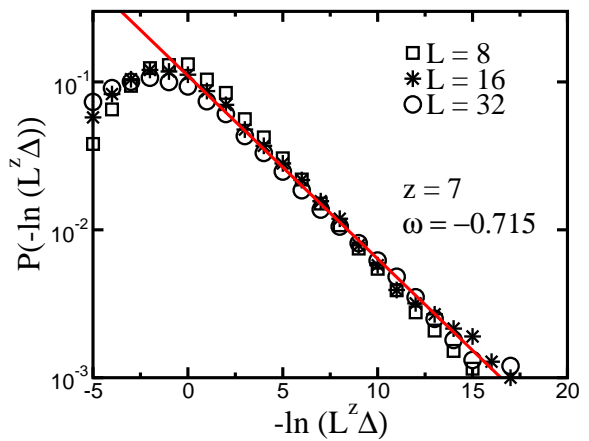}} \par}
 \caption{
 \label{fig:2dbi-gap-sc}
 A finite size scaling plot of the distribution of the logarithm of
 the energy gap for the double-layer antiferromagnet with a bilayer
 coupling $g=1$, bond randomness $D=8$ and dimer dilution
 concentration $p=0.125$. The dynamical exponent $z$ and the slop
 $(-1-\omega)$ of small energy gaps agree well with the relation
 $z=2/(1+\omega)$, implying localized energy gaps.  }
\end{figure}

$\bullet$ Above the critical bilayer coupling $g > g_c(p,D)$, the
ground state is an effective singlet and in accordance with this the
low-temperature uniform susceptibility is characterized by a
non-universal exponent $\theta$. For $p<p_p$, there is a infinite
cluster and the low-energy physics is governed by rare finite regions
which are locally ordered.  The low-energy excitations connected to
these regions are thus expected to be localized, provided the bond
disorder is sufficiently strong. This is illustrated in
Fig.\ref{fig:2dbi-gap-sc} in which the scaling collapse of the energy
gap distribution is obtained for $z=z'$ in accordance with
Eq.(\ref{eq:distr}). For $p>p_p$, the connected spin clusters are
finite and isolated. Therefore the low-energy excitations are also
localized.

\section{Summary and discussion}
\label{sec:disc}

In this paper we have studied the effect of strong bond disorder on
the low-energy, long-distance properties of Heisenberg
antiferromagnetic layers and bilayers with site and dimer dilution.
In particular we are interested in the structure of the phase diagram,
the form of the critical singularities as well as the properties of
the Griffiths singularities.

In a single layer the order-disorder transition point is found to be
at the percolation threshold $p=p_p$, thus for $p<p_p$ the AF order
survives for any finite value of bond disorder strength $D$.  In
contrast to this, the AF order at the percolation cluster, which is
present for $D=0$, is destroyed by bond disorder. This information is
deduced from the decay of the average spin-spin correlation function
which has the same power-law form with a strong-$D$ independent
exponent $2 \beta/\nu$, but much smaller than the known exponent $2
\beta/\nu=10/48$ at $D=0$.  The dynamical exponent $z$ of the diluted
single layer is found to be a continuously increasing function of the
disorder $D$.  Here we note that in the limit of infinite $D$ the
fixed point becomes an infinite disorder fixed point with $z\to
\infty$ so that the RG method is expected to be asymptotically exact
with increasing $D$, as well supported by comparing with QMC results.

In the dimer diluted bilayer with $g>0$, weak disorder is found not to
modify the static critical exponent $\beta/\nu$ as well as the
dynamical exponent $z$, which are - within the error bars - the same
as one measures at the fixed point $B$. On the other hand for strong
bond disorder the critical bilayer coupling is reduced to a very small
value and both the static and the dynamical exponents are different
than for weak disorder. While the static exponent approaches a $D$
independent limiting value the dynamical exponent shows a linear $D$
dependence.

Considering the Griffiths singularities the low-energy fixed point of
the RG is found to depend on the specific form of the disorder. For
example, the non-diluted single layer ($g=p=0$) transforms into an
effective singlet, the diluted single layer ($g=0$, $0<p<p_p$) into a
large spin, whereas the dimer diluted bilayer also into an effective
singlet. In each cases the disorder induced dynamical exponent is
found $D$ dependent for sufficiently large $D$. For smaller values of
$D$ the true dynamical exponent is determined by quantum fluctuations,
so that in this region disorder can influence only the corrections to
scaling. The low energy excitations are found to be non-localized for
weak bond disorder as well as in the large-spin phase, and become
localized only for substantially large disorder.

Useful discussions with A.~Sandvik, S.~Wessel and A.~L\"auchli 
are gratefully acknowledged.
This work has been supported by a German-Hungarian
exchange program (DAAD-M\"OB), by the Hungarian National Research Fund
under grant No OTKA TO37323, TO48721, MO45596 and M36803. NL acknowledges 
NSERC of Canada for financial support and 
WestGrid for access to computational facilities.


\begin{thebibliography}{99}
\bibitem{NOTE} We have checked that in the temperature range presented here the results 
               for the $32\times 32$ lattice are size independent.
\end{thebibliography}
\end{document}